# Visualization of three-dimensional incompressible flows by quasi-two-dimensional divergence-free projections

**Alexander Yu. Gelfgat**

School of Mechanical Engineering, Faculty of Engineering, Tel-Aviv University, Ramat Aviv, Tel-Aviv, Israel, 69978, gelfgat@tau.ac.il

**Abstract**

A visualization of three-dimensional incompressible flows by divergence-free quasi-two-dimensional projections of velocity field on three coordinate planes is proposed. It is argued that such divergence-free projections satisfying all the velocity boundary conditions are unique for a given velocity field. It is shown that the projected fields and their vector potentials can be calculated using divergence-free Galerkin bases. Using natural convection flow in a laterally heated cube as an example, it is shown that the projections proposed allow for a better understanding of similarities and differences of three-dimensional flows and their two-dimensional likenesses. An arbitrary choice of projection planes is further illustrated by a lid-driven flow in a cube, where the lid moves parallel either to a sidewall or a diagonal plane.

## 1. Introduction

With the growth of available computer power, development of numerical methods and experimental techniques dealing with fully developed three-dimensional flows the importance of flow visualization becomes obvious. While two-dimensional flows can be easily described by streamline or vector plots, there is no commonly accepted methodology for representation of three-dimensional flows on a 2D plot. Streamlines cannot be defined for a general 3D flow. Other textbook techniques, such as streak lines, trajectories and arrow fields, are widely used but become unhelpful with increase of flow complicacy. Same can be said about plotting of isosurfaces and isolines of velocity or vorticity components, which produce beautiful pictures, however, do not allow one to find out velocity direction at a certain point. Basic and more advanced recent state-of-the-art visualization techniques are discussed in review papers [1-3] where reader is referred for the details. Here we develop another visualization technique, applicable only to incompressible flows, and related to the surface-based techniques discussed in [2]. Our technique considers projections of 3D velocity field onto coordinate planes and allows one to compute a set of surfaces to which the projected flow is tangent. Thus, the flow is visualized in all three sets of coordinate planes (surfaces). The choice of visualization coordinate system is arbitrary, so that the axes can be directed along "most interesting" directions, e.g. directions parallel and orthogonal to dominating velocity or vorticity.

The visualization of three-dimensional incompressible flows described below is based on divergence-free projections of a 3D velocity field on two-dimensional coordinate planes. Initially, this study was motivated by a need to visualize three-dimensional benchmark flows, which are direct extensions of well-known two-dimensional benchmarks, e.g., lid-driven cavity and convection in laterally heated rectangular cavities. Thus, we seek for a visualization that is capable to show clearly both similarities and differences of flows considered in 2D and 3D formulations. It seems, however, that the technique proposed can have significantly wider area of applications.

Consider a given velocity field, which can be a result of computation or experimental measurement. Note, that modern means of flow measurement, like PIV and PTV, allow one to measure three velocity components on quite representative grids, which leads to the same problem of visualization of results. Here we observe that a three-dimensional divergence-free velocity field can be represented as a superposition of two vector fields that describe the motion in two sets of coordinate planes, say ($x$-$z$) and ($y$-$z$), without a need to consider the ($x$- *y*) planes. These fields allow for definition of vector potential of velocity, whose two independent components have properties of two-dimensional stream function. The two parts of velocity field are tangent to isosurfaces of the vector potential components, which allows one to visualize the flow in two sets of orthogonal coordinate planes. This approach, however, does not allow one to preserve the velocity boundary conditions in each of the fields separately, so that some of the boundary conditions are satisfied only after both fields are superimposed. The latter is not good for the visualization purposes. We argue further, that it is possible to define divergence-free projections of the flow on the three sets of coordinate planes, so that (i) the projections are unique, (ii) each projection is described by a single component of its vector potential, and (iii) the projection vectors are tangent to isosurfaces of the corresponding non-zero vector potential component. This allows us to visualize the flow in three orthogonal sets of coordinate planes. In particular, it helps to understand how the three-dimensional model flows differ from their two-dimensional likenesses. To calculate the projections we offer to use divergent-free Galerkin bases, on which the initial flow can be orthogonally projected.

For a representative example, we choose convection in a laterally heated square cavity with perfectly thermally insulated horizontal boundaries, and the corresponding three-dimensional extension, i.e., convection in a laterally heated cube with perfectly insulated horizontal and spanwise boundaries. The most representative solutions for steady states in these model flows can be found in [4] for the 2D benchmark, and in [5,6] for the 3D one. In these benchmarks the pressure $p$, velocity $\boldsymbol{v} = (u, v, w)$ and temperature $T$ are obtained as a solution of Boussinesq equations

$$\frac{\partial T}{\partial t} + (\boldsymbol{v} \cdot \nabla)T = \Delta T \tag{1}$$

$$\frac{\partial \boldsymbol{v}}{\partial t} + (\boldsymbol{v} \cdot \nabla)\boldsymbol{v} = -\nabla p + Pr\Delta \boldsymbol{v} + RaPrT\boldsymbol{e}_z \tag{2}$$

$$div\,\boldsymbol{v} = 0 \tag{3}$$

defined in a square $0 \leq x, z \leq 1$ or in a cube $0 \leq x, y, z \leq 1$, with the no-slip boundary conditions on all the boundaries. The boundaries $x = 0,1$ are isothermal and all the other boundaries are thermally insulated, which in the dimensionless formulation reads

$$T(x=0) = 1, \quad T(x=1) = 0, \quad \left(\frac{\partial T}{\partial y}\right)_{y=0,1} = 0, \quad \left(\frac{\partial T}{\partial z}\right)_{z=0,1} = 0. \tag{4}$$

$Ra$ and $Pr$ are the Rayleigh and Prandtl numbers. The reader is referred to the above cited papers for more details. Here we focus only on visualization of solutions of 3D problem and comparison with the corresponding 2D flows. All the flows reported below are calculated on $100^2$ and $100^3$ stretched finite volume grids, which is accurate enough for present visualization purposes (for convergence studies see also [7]).

Apparently, the 2D flow $\boldsymbol{v} = (u, 0, w)$ is best visualized by the streamlines, which are the isolines of the stream function $\psi$ defined as $u = \frac{\partial \psi}{\partial z}, w = -\frac{\partial \psi}{\partial x}$. In each point the velocity vector is tangent to a streamline passing through the same point, so that plot of streamlines and schematic indication of the flow direction is sufficient to visualize a two-dimensional flow. This is illustrated in Fig. 1, where streamlines of flows calculated for $Pr = 0.71$, and $Ra$ varied from $10^3$ to $10^8$ are shown. Note how the streamline patterns complicate with the increase of Rayleigh number. Our further purpose is to visualize three-dimensional flows at the same Rayleigh numbers, so that it will be possible to see similarities and differences of 2D and 3D flows.

2. **Preliminary considerations**

We consider an incompressible flow in a rectangular box $0 \leq x \leq X, 0 \leq y \leq Y, 0 \leq z \leq Z$, satisfying the no-slip conditions on all boundaries. The continuity equation $\partial u/\partial x + \partial v/\partial y + \partial w/\partial z = 0$ makes one velocity component dependent on two others, so that to describe the velocity field we need two scalar three-dimensional functions, while the third one can be found via continuity. This observation allows us to decompose the velocity field in the following way

$$\boldsymbol{v} = \begin{bmatrix} u \\ v \\ w \end{bmatrix} = \begin{bmatrix} u \\ 0 \\ w_1 \end{bmatrix} + \begin{bmatrix} 0 \\ v \\ w_2 \end{bmatrix} \quad, \quad w_1 = -\int_0^z \frac{\partial u}{\partial x} dz, \quad w_2 = -\int_0^z \frac{\partial v}{\partial y} dz \qquad (5)$$

This decomposition shows that the div-free velocity field can be represented as superposition of two fields having components only in the (*x,z*) or (*y,z*) planes. Moreover, we can easily define the vector potential of velocity field as

$$\boldsymbol{\Psi} = \begin{bmatrix} \Psi_x \\ \Psi_y \\ 0 \end{bmatrix} = \begin{bmatrix} \int_0^z v dz \\ -\int_0^z u dz \\ 0 \end{bmatrix} \quad, \quad \boldsymbol{v} = rot \boldsymbol{\Psi} \qquad (6)$$

Thus, $\boldsymbol{\Psi}$ is the vector potential of velocity field $\boldsymbol{v}$, and its two non-zero components have properties of the stream function:

$$u = -\frac{\partial \Psi_y}{\partial z}, \quad w_1 = \frac{\partial \Psi_y}{\partial x}; \quad v = \frac{\partial \Psi_x}{\partial z}, \quad w_2 = -\frac{\partial \Psi_x}{\partial y}. \qquad (7)$$

This means, in particular, that vectors of the two components of decomposition (5), i.e., $(u, 0, w_1)$ and $(0, v, w_2)$, are tangent to isosurfaces of $\Psi_y$ and $\Psi_x$, and the vectors are located in the planes (*x-z*) and (*y-z*), respectively. Thus, it seems that the isosurfaces of $\Psi_y$ and $\Psi_x$, which can be easily calculated from numerical or experimental (e.g., PIV) data, can be a good means for visualization of the velocity field. Unfortunately, there is a drawback, which can make such a visualization meaningless. Namely, only the sum of vectors $w_1$ and $w_2$, calculated via the integrals in (5), vanish at $z = Z$, while each vector separately does not. Thus, visualization of flow via the decomposition (5) in a straight-forward way will result in two fields that violate no-penetration boundary conditions at one of the boundaries, which would make the whole result quite meaningless. The latter is illustrated in Fig. 2, where isosurfaces of the two components of vector potentials are superimposed with the $(u, 0, w_1)$ and $(0, v, w_2)$ vectors. It is clearly seen that the vector arrows are tangent to the isosurfaces, however the velocities $w_1$ and $w_2$ do not vanish at the upper boundary. Moreover, the choice of integration boundaries in (5) is arbitrary, so that the whole decomposition (5) is not unique. Clearly, one would prefer to visualize unique properties of the flow rather than non-unique ones.

To define unique flow properties similar to those shown in Fig. 2 we observe that decomposition (5) can be interpreted as representation of a three-dimensional divergence-free vector into two divergence free vector fields located in orthogonal coordinate planes, i.e., having only two non-zero components. Consider a vector built from only two components of the initial field, say $\boldsymbol{u} = (u, v, 0)$. It is located in the $(x, y, z = const)$ planes, satisfies all the boundary conditions for $u$ and $v$, however, is not divergence-free. We can apply the Helmholtz-Leray decomposition [8] that decomposes this vector into solenoidal and potential part,

$$\boldsymbol{u} = \nabla \varphi + \hat{\boldsymbol{u}}, \quad \nabla \cdot \hat{\boldsymbol{u}} = 0 \tag{8}$$

As is shown in [8], together with the boundary conditions

$$\hat{\boldsymbol{u}} \cdot \boldsymbol{n} = \boldsymbol{0}, \quad \text{and} \quad \frac{\partial \varphi}{\partial n} = \boldsymbol{u} \cdot \boldsymbol{n} \tag{9}$$

where $\boldsymbol{n}$ is a normal to the boundary, the decomposition (8) is unique. For the following, we consider (8) in the $(x, y, z = const)$ planes and seek for a decomposition of $\boldsymbol{u} = (u, v, 0)$ in a $(x, y, z = const)$ plane

$$\boldsymbol{u} = \nabla_{(x,y)} \varphi + \hat{\boldsymbol{u}}, \quad \nabla_{(x,y)} \cdot \hat{\boldsymbol{u}} = 0, \quad \nabla_{(x,y)} = \boldsymbol{e}_x \frac{\partial}{\partial x} + \boldsymbol{e}_y \frac{\partial}{\partial y}. \tag{10}$$

We represent the divergent-free two-dimensional vector $\hat{\boldsymbol{u}} = (\hat{u}, \hat{v}, 0)$ as

$$\hat{\boldsymbol{u}} = rot \boldsymbol{\Psi}, \quad \boldsymbol{\Psi} = (0, 0, \Psi_z) \quad \Rightarrow \quad \hat{u} = \frac{\partial \Psi_z}{\partial y}, \quad \hat{v} = -\frac{\partial \Psi_z}{\partial x} \tag{11}$$

which yields for the *z*-component of $rot\mathbf{u}$:

$$\mathbf{e}_z \cdot rot\mathbf{u} = \mathbf{e}_z \cdot rot\hat{\mathbf{u}} = \mathbf{e}_z \cdot rotrot\mathbf{\Psi} = -\mathbf{e}_z \cdot \Delta\mathbf{\Psi} = -\Delta\Psi_z \qquad (12)$$

This shows that $\Psi_z$ is an analog of the two-dimensional stream function, so that in each plane $(x, y, z = const)$ vector $\hat{\mathbf{u}}$ is tangent to an isoline of $\Psi_z$. To satisfy the no-slip boundary conditions for $\hat{u}$ and $\hat{v}$, $\Psi_z$ and its normal derivative must vanish on the boundary, which makes the definition of both $\Psi_z$ and $\hat{\mathbf{u}}$ unique. Note, that contrarily to the boundary conditions (9), to make vector $\hat{\mathbf{u}}$ in the decomposition (10) unique and satisfying all the boundary conditions of $\mathbf{u}$, we do not need to define any boundary conditions for the scalar potential $\varphi$.

To conclude, the resulting solenoidal part $\hat{\mathbf{u}}$ of vector $\mathbf{u} = (u, v, 0)$ (i) is unique, (ii) is defined by a single non-zero *z*-component $\Psi_z$ of its vector potential, and (iii) in each $(x, y, z = const)$ plane vectors of $\hat{\mathbf{u}}$ are tangent to the isosurfaces of $\Psi_z$. Defining same solenoidal fields for two other sets of coordinate planes we arrive to three quasi-two-dimensional divergent free projections of the initial velocity field. Each projection is described by a single scalar three-dimensional function, which, in fact, is a single non-zero component of the corresponding vector potential.

In the following we use the three above quasi-two-dimensional divergence-free projections for visualization of convective flow in a laterally heated cube, and offer a way to calculate them. In particular, to compare a three-dimensional result with the corresponding two-dimensional one, we need to compare one of the projections. Thus, if the 2D convective flow was considered in the plane $(x, z)$, we compare it with the corresponding projections of the 3D flow on the $(x, y = const, z)$ planes, which are tangent to isosurfaces of the non-zero *y*-component of the corresponding vector potential.

### 3. Numerical realization

A direct numerical implementation of the Helmholtz-Leray decomposition to an arbitrary velocity field is known in CFD as Chorin projection. This procedure is well-known, uses the boundary conditions (9), but does not preserve all the velocity boundary conditions. Therefore, it is not applicable for our purposes. Alternatively, we propose orthogonal projections of the initial velocity field on divergence-free Galerkin bases used previously for computations of different two-dimensional flows.

Divergence-free basis functions that satisfy all the linear homogeneous boundary conditions were introduced in [9] for two-dimensional flows and were then extended in [10,11] to three-dimensional cases. To make further numerical process clear we briefly describe these bases below. The bases are built from shifted Chebyshev polynomials of the 1st and 2nd kind, $T_n(x)$ and $U_n(x)$, that are defined as

$$T_n(x) = cos[narccos(2x-1)], \quad U_n(x) = \frac{sin[(n+1)arccos(2x-1)]}{sin[arccos(2x-1)]} \quad (13)$$

and are connected via derivative of $T_n(x)$ as $T'_n(x) = 2nU_{n-1}(x)$. Each of system of polynomials, either $T_n(x)$ or $U_n(x)$, form basis in the space of continuous functions defined on the interval $0 \leq x \leq 1$. It is easy to see that vectors

$$\hat{\boldsymbol{q}}_{ij}^{2D} = \begin{bmatrix} \frac{X}{2i} T_i\left(\frac{x}{X}\right) U_{j-1}\left(\frac{y}{Y}\right) \\ -\frac{Y}{2j} U_{i-1}\left(\frac{x}{X}\right) T_j\left(\frac{y}{Y}\right) \end{bmatrix} \quad (14)$$

form a divergent-free basis in the space of divergent-free functions defined on a rectangle $0 \leq x \leq X, 0 \leq y \leq Y$. Assume that a two-dimensional problem is defined with two linear and homogeneous boundary conditions for velocity at each boundary, e.g., the no-slip conditions. This yields four boundary conditions in either x- or y-direction for the two velocity components. To satisfy the boundary conditions we extend components of the vectors (14) into linear superpositions as

$$\boldsymbol{q}_{ij}^{2D} = \begin{bmatrix} \frac{X}{2} \sum_{l=0}^{4} \frac{a_{il}}{(i+l)} T_{i+l}\left(\frac{x}{X}\right) \sum_{m=0}^{4} b_{jm} U_{j+m-1}\left(\frac{y}{Y}\right) \\ -\frac{Y}{2} \sum_{l=0}^{4} a_{il} U_{i+l-1}\left(\frac{x}{X}\right) \sum_{m=0}^{4} \frac{b_{jm}}{(j+m)} T_{j+m}\left(\frac{y}{Y}\right) \end{bmatrix} \quad (15)$$

For each $i$ a substitution of (15) into the boundary conditions yields four linear homogeneous equations for five coefficients $a_{il}, l = 0,1,2,3,4$. Fixing $a_{i0} = 1$, allows one to define all the other coefficients, whose dependence on $i$ and $l$ can be derived analytically. The coefficients $b_{jm}$ are evaluated in the same way. Expressions for these coefficients for the no-slip boundary conditions can be found in [9,11]. Since the basis functions $\boldsymbol{q}_{ij}^{2D}$ are divergence-free in the plane $(x,y)$, $\partial(\boldsymbol{q}_{ij}^{2D})_x/\partial x + \partial(\boldsymbol{q}_{ij}^{2D})_y/\partial y = 0$, and satisfy the non-penetration conditions through all the boundaries $x = 0, X$ and $y = 0, Y$, they are orthogonal to every two-dimensional potential vector field, i.e.,

$$\int_0^Y \int_0^X \nabla_{2D} p \cdot \boldsymbol{q}_{ij}^{2D} dxdy = \int_0^Y \int_0^X \left(\frac{\partial p}{\partial x} \boldsymbol{e}_x + \frac{\partial p}{\partial y} \boldsymbol{e}_y\right) \cdot \boldsymbol{q}_{ij}^{2D} dxdy = 0, \quad (16)$$

which is an important point for further evaluations.

For extension of the two-dimensional basis to the three dimensional case we recall that for divergence-free vector field we have to define independent three-dimensional bases for two components only. Representing the flow in the form (6) and using the same idea as in the two-dimensional basis (15) we arrive to a set of three-dimensional basis functions formed from two following subsets

$$\boldsymbol{q}_{ijk}^{(y)}(x,y,z) = \begin{bmatrix} \frac{X}{2}\sum_{l=0}^{4}\frac{\hat{a}_{il}}{(i+l)}T_{i+l}\left(\frac{x}{X}\right)\sum_{m=0}^{4}\hat{b}_{jm}T_{j+m}\left(\frac{y}{Y}\right)\sum_{n=0}^{4}\hat{c}_{kn}U_{k+n-1}\left(\frac{z}{Z}\right) \\ 0 \\ -\frac{Z}{2}\sum_{l=0}^{4}\hat{a}_{il}U_{i+l-1}\left(\frac{x}{X}\right)\sum_{m=0}^{4}\hat{b}_{jm}T_{j+m}\left(\frac{y}{Y}\right)\sum_{n=0}^{4}\frac{\hat{c}_{kn}}{(k+n)}T_{k+n}\left(\frac{z}{Z}\right) \end{bmatrix} \quad (17)$$

$$\boldsymbol{q}_{ijk}^{(x)}(x,y,z) = \begin{bmatrix} 0 \\ \frac{Y}{2}\sum_{l=0}^{4}\tilde{a}_{il}T_{i+l}\left(\frac{x}{X}\right)\sum_{m=0}^{4}\frac{\tilde{b}_{jm}}{(j+m)}T_{j+m}\left(\frac{y}{Y}\right)\sum_{n=0}^{4}\tilde{c}_{kn}U_{k+n-1}\left(\frac{z}{Z}\right) \\ -\frac{Z}{2}\sum_{l=0}^{4}\tilde{a}_{il}T_{i+l}\left(\frac{x}{X}\right)\sum_{m=0}^{4}\tilde{b}_{jm}U_{j+m-1}\left(\frac{y}{Y}\right)\sum_{n=0}^{4}\frac{\tilde{c}_{kn}}{(k+n)}T_{k+n}\left(\frac{z}{Z}\right) \end{bmatrix} \quad (18)$$

Where coefficients $\hat{a}_{il}, \hat{b}_{jm}, \tilde{c}_{kn}, \tilde{a}_{il}, \tilde{b}_{jm}, \tilde{c}_{kn}$ are defined from the boundary conditions. Their expressions for no-slip boundary conditions are given in [11]. The velocity field is approximated as a truncated series

$$\hat{\boldsymbol{u}} \approx \hat{\boldsymbol{u}}^{(x)} + \hat{\boldsymbol{u}}^{(y)}, \quad \hat{\boldsymbol{u}}^{(x)} = \sum_{i=0}^{N_x}\sum_{j=0}^{N_y}\sum_{k=0}^{N_z}B_{ijk}\boldsymbol{q}_{ijk}^{(x)}, \quad \hat{\boldsymbol{u}}^{(y)} = \sum_{i=0}^{N_x}\sum_{j=0}^{N_y}\sum_{k=0}^{N_z}A_{ijk}\boldsymbol{q}_{ijk}^{(y)} \quad (19)$$

Here one must be cautious with the boundary conditions in *z*-direction since, as it was explained above, the two parts of representation (6) satisfy the boundary condition for *w* at $z = Z$ only as a sum. Therefore we must exclude this condition from definition of basis functions (17) and (18) and set $\hat{c}_{k4} = \tilde{c}_{k4} = 0$. The corresponding boundary condition should be included in the resulting system of equations for $A_{ijk}$ and $B_{ijk}$ as an additional algebraic constraint. Note that this fact was overlooked in [10] that could lead to missing of some important three-dimensional Rayleigh-Bénard modes. On the other hand, comparison of 3D basis functions (16) and (17) with the 2D ones (15) shows that with all the boundary conditions included, the functions $\boldsymbol{q}_{ijk}^{(y)}(x,y,z)$ and $\boldsymbol{q}_{ijk}^{(x)}(x,y,z)$ form the complete two-dimensional bases in the $(x, y = const, z)$ and $(x = const, y, z)$ planes, respectively. The coefficients $\hat{b}_{jm}$ and $\tilde{a}_{il}$ are used to satisfy the boundary conditions in the third direction. For the basis in the $(x, y, z = const)$ planes we add

$$\boldsymbol{q}_{ijk}^{(z)}(x,y,z) = \begin{bmatrix} \frac{X}{2}\sum_{l=0}^{4}\frac{\bar{a}_{il}}{(i+l)}T_{i+l}\left(\frac{x}{X}\right)\sum_{m=0}^{4}\bar{b}_{jm}U_{j+m-1}\left(\frac{y}{Y}\right)\sum_{n=0}^{4}\bar{c}_{kn}T_{j+n}\left(\frac{z}{Z}\right) \\ \frac{-Y}{2}\sum_{l=0}^{4}\bar{a}_{il}U_{i+l-1}\left(\frac{x}{X}\right)\sum_{m=0}^{4}\frac{\bar{b}_{jm}}{(j+m)}T_{j+m}\left(\frac{y}{Y}\right)\sum_{n=0}^{4}\bar{c}_{kn}T_{j+n}\left(\frac{z}{Z}\right) \\ 0 \end{bmatrix} \quad (20)$$

Now, we define an inner product as

$$\langle \boldsymbol{u}, \boldsymbol{v} \rangle = \int_V \boldsymbol{u} \cdot \boldsymbol{v}\, dV \tag{21}$$

and compute projections of the initial velocity vector on each of the three basis systems separately. Together with the vectors $\hat{\boldsymbol{u}}^{(x)}$ and $\hat{\boldsymbol{u}}^{(y)}$ defined in Eq. (19) we obtain also vector

$$\hat{\boldsymbol{u}}^{(z)} = \sum_{i=0}^{N_x}\sum_{j=0}^{N_y}\sum_{k=0}^{N_z} C_{ijk}\boldsymbol{q}_{ijk}^{(z)} \tag{22}$$

Since the basis vectors $\boldsymbol{q}_{ijk}^{(x)}$, $\boldsymbol{q}_{ijk}^{(y)}$ and $\boldsymbol{q}_{ijk}^{(z)}$ satisfy all the boundary conditions and are divergence-free not only in the 3D space, but also into the corresponding coordinate planes, the potential parts of projections on these planes are excluded by (16), and resulting vectors $\hat{\boldsymbol{u}}^{(x)}, \hat{\boldsymbol{u}}^{(y)}, \hat{\boldsymbol{u}}^{(z)}$ satisfy the boundary conditions and are divergence free in the planes they are located. Therefore, they approximate the quasi-two-dimensional divergent-free projection vectors we are looking for. Note, however, that the superposition $\hat{\boldsymbol{u}}^{(x)} + \hat{\boldsymbol{u}}^{(y)} + \hat{\boldsymbol{u}}^{(z)}$ does not approximate the initial vector $\boldsymbol{v}$. To complete the visualization we have to derive the corresponding approximation of vector potentials. The vector potential of each of $\hat{\boldsymbol{u}}^{(x)}, \hat{\boldsymbol{u}}^{(y)}, \hat{\boldsymbol{u}}^{(z)}$ has only one non-zero component, as is defined below

$$\hat{\boldsymbol{u}}^{(x)} = rot\boldsymbol{\Psi}^{(x)}, \quad \boldsymbol{\Psi}^{(x)} = \left(\Psi_x^{(x)}, 0, 0\right), \quad \Psi_x^{(x)} \approx \sum_{i=0}^{N_x}\sum_{j=0}^{N_y}\sum_{k=0}^{N_z} B_{ijk}\varphi_{ijk}^{(x)},$$

$$\hat{\boldsymbol{u}}^{(y)} = rot\boldsymbol{\Psi}^{(y)}, \quad \boldsymbol{\Psi}^{(y)} = \left(0, \Psi_y^{(y)}, 0\right), \quad \Psi_y^{(y)} \approx \sum_{i=0}^{N_x}\sum_{j=0}^{N_y}\sum_{k=0}^{N_z} A_{ijk}\varphi_{ijk}^{(y)}, \tag{23}$$

$$\hat{\boldsymbol{u}}^{(z)} = rot\boldsymbol{\Psi}^{(z)}, \quad \boldsymbol{\Psi}^{(z)} = \left(0, 0, \Psi_z^{(z)}\right), \quad \Psi_z^{(z)} \approx \sum_{i=0}^{N_x}\sum_{j=0}^{N_y}\sum_{k=0}^{N_z} C_{ijk}\varphi_{ijk}^{(z)}$$

where

$$\varphi_{ijk}^{(x)}(x,y,z) = -\sum_{l=0}^{4}\tilde{a}_{il}T_{i+l}\left(\frac{x}{X}\right)\sum_{m=0}^{4}\frac{\tilde{b}_{jm}}{(j+m)}T_{j+m}\left(\frac{y}{Y}\right)\sum_{n=0}^{4}\frac{\tilde{c}_{kn}}{(k+n)}T_{k+m}\left(\frac{z}{Z}\right)$$

$$\varphi_{ijk}^{(y)}(x,y,z) = \sum_{l=0}^{4}\frac{\hat{a}_{il}}{(i+l)}T_{i+l}\left(\frac{x}{X}\right)\sum_{m=0}^{4}\hat{b}_{jm}T_{j+m}\left(\frac{y}{Y}\right)\sum_{n=0}^{4}\frac{\hat{c}_{kn}}{(k+n)}T_{k+m}\left(\frac{z}{Z}\right) \tag{24}$$

$$\varphi_{ijk}^{(z)}(x,y,z) = \sum_{l=0}^{4}\frac{\bar{a}_{il}}{(i+l)}T_{i+l}\left(\frac{x}{X}\right)\sum_{m=0}^{4}\frac{\bar{b}_{jm}}{(j+m)}T_{j+m}\left(\frac{y}{Y}\right)\sum_{n=0}^{4}\bar{c}_{kn}T_{k+m}\left(\frac{z}{Z}\right)$$

As stated above, the vectors $\hat{u}^{(x)}, \hat{u}^{(y)}, \hat{u}^{(z)}$ are tangent to the isosurfaces of $\Psi_x^{(x)}$, $\Psi_y^{(y)}$ and $\Psi_z^{(z)}$, respectively.

## 4. Visualization results

We start from the flow at $Ra = 10^3$, that has the simplest pattern (Fig. 3). Figure 3a shows two trajectories starting in points (0.1,0.1,0.1) and (0.9,0.9,0.9). The trajectories are colored according to the values of temperature they pass, so that it is clearly seen that the fluid rises near the hot wall and descends near the cold one. Looking only at the trajectories, one can mistakenly conclude that convective circulation weakens toward the center plane $y = 0.5$. Frames 3b-3d show that this impression is misleading. In these frames we plot three vector potentials defined in Eqs. (23), together with the divergent-free velocity projections (shown by arrows) on the corresponding coordinate planes. First, it is clearly seen that the projection vectors are tangent to the isolines of the vector potentials. Then we observe that projections on the $y = const.$ planes (Fig. 3b) represent the simple convective two-dimensional circulation shown in Fig. 1a. Contrarily to the impression of Fig. 3a, the circulations in $(x, z)$ are almost *y*-independent near the center plane $y = 0.5$ and steeply decay near the boundaries $y = 0$ and $y = 1$. The three-dimensional effects are rather clearly seen from the two remaining frames. The flow contains two pairs of diagonally symmetric rolls in the $(y, z)$ planes (Fig. 3c), and two other diagonally symmetric rolls in the $(x, y)$ planes (Fig. 3d). Motion along these rolls deforms trajectories shown in Fig. 3a.

It is intuitively clear that the motion in the frames of Fig. 3c and 3d is noticeably weaker than that in Fig. 3b. For the 2D flows the integral intensity of convective circulation can be estimated by the maximal value of the stream function. Similarly, here we can estimate the intensity of motion in two-dimensional planes by maximal values of the corresponding vector potential. Since these values can be used also for comparison of results obtained by different methods we report all of them, together with their locations, in Table 1. As expected, we observe that at $Ra = 10^3$ the maximal value of $\Psi_y^{(y)}$ is larger than that of two other potentials in almost an order of magnitude. With the increase of Rayleigh number the ratio of maximal values of $\Psi_x^{(x)}$, $\Psi_z^{(z)}$ and $\Psi_y^{(y)}$ grows reaching approximately one half at $Ra = 10^7$, which indicates on the growing importance of motion in the third direction.

Figures 4-6 illustrate flows at $Ra = 10^5$, $10^7$ and $10^8$, respectively, in the same way as in Fig. 3. It is seen that the isosurfaces of $\Psi_y^{(y)}$ resemble the shapes of two-dimensional streamlines (Fig. 1) rather closely. At the same time we see that the "three-dimensional additions" to the flow, represented by $\Psi_x^{(x)}$ and $\Psi_z^{(z)}$, remain located near the no-slip boundaries and are weak in the central region of the cavity. This means, in particular, that spanwise directed motion in the midplane $y = 0.5$ is weak, which justifies use of two-dimensional model for description of the main convective circulation.

To show how the isosurfaces of $\Psi_y^{(y)}$ represent patterns of two-dimensional flow we show their several isosurfaces in Fig. 7 and isolines in the center plane $y = 0.5$ in Fig. 8. The isosurfaces of $\Psi_y^{(y)}$ in Fig. 7 shows pattern of the main convective circulation in the $(x, y = const, z)$ planes. The isolines in Fig. 8 can be directly compared with the streamlines shown in Fig. 1. This comparison should be accompanied with the comparison of the maximal values of the stream functions of Fig. 1 and the maximal values of $\Psi_y^{(y)}$, all shown in the figures. We observe that the patterns in Figs. 1 and 8 remain similar, however the similarity diminishes with the increase of $Ra$. The maximal values of $\Psi_y^{(y)}$ for $Ra \leq 10^5$ are smaller than that of the stream function, which can be easily explained by additional friction losses due to the spanwise boundaries added to the three-dimensional formulation. At $Ra \geq 10^6$ we observe that together with the deviation of the isolines pattern from the 2D one, the maximal values of $\Psi_y^{(y)}$ become larger than those of the two-dimensional stream function. Ensuring, that this is not an effect of truncation in the sums (23), we explain this by strong three-dimensional effects, in which motion along the *y*-axis start to affect the motion in the $(x, y = const, z)$ planes.

As an example of arbitrary choice of projection planes we consider another well-known benchmark problem of flow in a lid-driven cubic cavity. We consider it in two different formulations: a classical configuration where the lid moves parallel to a side wall, and a modified configuration with the lid moving along the diagonal of the upper boundary [12]. Obviously, three-dimensional effects are significantly stronger in the second case. Both flows are depicted in Figs. 9 and 10 in the same way as convective flows were represented above. Comparing the flow pattern shown in Fig. 9, one can see clear similarities with the well-known two-dimensional flow in a lid-driven cavity. The main vortex and reverse recirculation in the lower corner are clearly seen in Fig, 9b. Figures 9c and 9d show additional three-dimensional recirculations in the $(x, y, z = const)$ and $(x = const, y, z)$ planes. The same representation of the second configuration in Fig. 10 exhibits similar patterns of $\Psi_x^{(x)}$ and $\Psi_y^{(y)}$ together with the similar patterns of corresponding projection vectors. This is an obvious consequence of the problem configuration, where main motion is located in the diagonal plane and the planes parallel to it. To illustrate motion in these planes we project the flow on planes orthogonal to the diagonal plane (or parallel to the second diagonal plane). The result is shown in Fig. 11. The isosurfaces belong to the corresponding vector potential, so that the divergence-free projection of velocity on the diagonal and parallel planes is tangent to these and other isosurfaces. Arrows in the diagonal plane depict this projection and illustrate the main vortex, as well as small recirculation vortices in lower corners. It is seen that the arrows are tangent to both isosurfaces.

## 5. Conclusions

We proposed to visualize three-dimensional incompressible flows by divergence-free projections of velocity field on three coordinate planes. We presented the arguments showing that such a representation allows, in particular, for a better understanding of similarities and differences between three-dimensional benchmark flow models and their two-dimensional counter parts. We argued also that the choice of projection planes is arbitrary, so that they can be fitted to the flow pattern.

To approximate the divergence-free projections numerically we calculated orthogonal projections on divergence-free Galerkin velocity bases. Obviously, there are other ways of doing that, among which we can mention inverse of the Stokes operator discussed in [13]. We believe also that the proposed method of visualization is suitable for a significantly wider class of incompressible flows, and can be applied not only to numerical, but also to experimental data.

**Acknowledgement**

This work was supported by the LinkSCEEM-2 project, funded by the European Commission under the 7th Framework Program through Capacities Research Infrastructure, INFRA-2010-1.2.3 Virtual Research Communities, Combination of Collaborative Project and Coordination and Support Actions (CP-CSA) under grant agreement no RI-261600.

**Figure captions**

Fig. 1. Streamlines of two-dimensional buoyancy convection flow in a laterally heated square cavity at *Pr*=0.71 and different Rayleigh numbers. The direction of main circulation is clockwise.

Fig. 2. Calculation for *Ra*=$10^3$. Vector potentials $\Psi_y$ and $\Psi_x$ defined in Eq. (6) superposed with the vector fields $(u, 0, w_1)$ and $(0, v, w_2)$. . (a) max$\Psi_y = 1.6$, isosurface for $\Psi_y$=0.375; (b) max$|\Psi_x| = 0.075$, isosurfaces for $\Psi_x = \pm 0.015$.

Fig. 3. Visualization of a three-dimensional flow at *Ra*=$10^3$. (a) Two flow trajectories starting at the points (0.1,0.1,0.1) and (0.9,0.9,0.9). The trajectories are colored due to the temperature values at the points they pass. The temperature color map is shown aside. (b), (c), (d) Isosurfaces of $\Psi_y^{(y)}$, $\Psi_x^{(x)}$ and $\Psi_z^{(z)}$ superimposed with the vector plots of the fields $\hat{\boldsymbol{u}}^{(y)}, \hat{\boldsymbol{u}}^{(x)}$ and $\hat{\boldsymbol{u}}^{(z)}$, respectively. The isosurfaces are plotted for $\Psi_y^{(y)} = 0.375$, $\Psi_x^{(x)} = \pm 0.088$, and $\Psi_z^{(z)} = \pm 0.11$.

Fig. 4. Visualization of a three-dimensional flow at *Ra*=$10^5$. (a) Two flow trajectories starting at the points (0.1,0.1,0.1) (0.9,0.9,0.9) and (0.4,0.5,0.5). The trajectories are colored due to the temperature values at the points they pass. The temperature color map is shown aside. (b), (c), (d) Isosurfaces of $\Psi_y^{(y)}$, $\Psi_x^{(x)}$ and $\Psi_z^{(z)}$ superimposed with the vector plots of the fields $\hat{\boldsymbol{u}}^{(y)}, \hat{\boldsymbol{u}}^{(x)}$ and $\hat{\boldsymbol{u}}^{(z)}$, respectively. The isosurfaces are plotted for $\Psi_y^{(y)} = 5.63$, $\Psi_x^{(x)} = \pm 0.64$, and $\Psi_z^{(z)} = \pm 0.27$.

Fig. 5. Visualization of a three-dimensional flow at *Ra*=$10^7$. (a) Two flow trajectories starting at the points (0.1,0.1,0.1), (0.9,0.9,0.9) and (0.1,0.5,0.5). The trajectories are colored due to the temperature values at the points they pass. The temperature color map is shown aside. (b), (c), (d) Isosurfaces of $\Psi_y^{(y)}$, $\Psi_x^{(x)}$ and $\Psi_z^{(z)}$ superimposed with the vector plots of the fields $\boldsymbol{v}^{(y)}, \boldsymbol{v}^{(x)}$ and $\boldsymbol{v}^{(z)}$, respectively. The isosurfaces are plotted for $\Psi_y^{(y)} = 28.5$, $\Psi_x^{(x)} = \pm 2.6$, and $\Psi_z^{(z)} = \pm 3.0$

Fig. 6. Visualization of a three-dimensional flow at *Ra*=$10^8$. (a) Two flow trajectories starting at the points (0.1,0.1,0.1) (0.9,0.9,0.9) and (0.1,0.5,0.5). The trajectories are colored due to the temperature values at the points they pass. The temperature color map is shown aside. (b), (c), (d) Isosurfaces of $\Psi_y^{(y)}$, $\Psi_x^{(x)}$ and $\Psi_z^{(z)}$ superimposed with the vector plots of the fields $\boldsymbol{v}^{(y)}, \boldsymbol{v}^{(x)}$ and $\boldsymbol{v}^{(z)}$, respectively. The isosurfaces are plotted for $\Psi_y^{(y)} = 66.0$, $\Psi_x^{(x)} = \pm 7.0$, and $\Psi_z^{(z)} = \pm 11.0$.

Fig. 7. Isosurfaces of $\Psi_y^{(y)}$ at different Rayleigh numbers. The isosurfaces are plotted at levels (a) 0.75, 5.6, 12.4; (b) 3.6, 15.4, 22.5; (c) 7.5, 24.4, 37.5; (d) 17.8, 47.5, 71.2 .

Fig. 8. Isolines of $\Psi_y^{(y)}$ in the midplane $y = 0.5$ at different Rayleigh numbers. The direction of main circulation is clockwise.

Fig. 9. Visualization of a three-dimensional flow in a lid-driven cubic cavity at $Re=10^3$. (a) Two flow trajectories starting at the points (0.4,0.4,0.9) and (0.6,0.6,0.9). The trajectories are colored due to values of spanwise velocity. (b), (c), (d) Isosurfaces of $\Psi_y^{(y)}$, $\Psi_x^{(x)}$ and $\Psi_z^{(z)}$ superimposed with the vector plots of the fields $\boldsymbol{v}^{(y)}, \boldsymbol{v}^{(x)}$ and $\boldsymbol{v}^{(z)}$, respectively. The isosurfaces are plotted for $\Psi_y^{(y)} = -0.0032$ and $+0.0008$; $\Psi_x^{(x)} = -0.0044$, and $+0.0033$; $\Psi_z^{(z)} = \pm 0.0028$.

Fig. 10. Visualization of a three-dimensional flow in a lid-driven cubic cavity with a lid moving along a diagonal, at $Re=10^3$. (a) Two flow trajectories starting at the points (0.1,0.1,0.9) and (0.9,0.9,0.9). The trajectories are colored due to values of vertical velocity. (b), (c), (d) Isosurfaces of $\Psi_y^{(y)}$, $\Psi_x^{(x)}$ and $\Psi_z^{(z)}$ superimposed with the vector plots of the fields $\boldsymbol{v}^{(y)}, \boldsymbol{v}^{(x)}$ and $\boldsymbol{v}^{(z)}$, respectively. The isosurfaces are plotted for $\Psi_y^{(y)} = -0.015$ and $+0.0015$; $\Psi_x^{(x)} = -0.015$, and $+0.0015$; $\Psi_z^{(z)} = \pm 0.0024$.

Fig. 11. Visualization of a three-dimensional flow in a lid-driven cubic cavity with a lid moving along a diagonal, at $Re=10^3$. Isosurfaces of vector potential of velocity projection on the diagonal planes, and the vector plot of the corresponding projected velocity field. The isosurfaces are plotted for the levels -0.017 and +0.004, while the minimal and maximal values of the calculated vector potential are -0.083 and +0.012 .

Table 1. Maximal values of the calculated vector potentials of the three velocity projections and their locations.

| Ra | $10^3$ | $10^4$ | $10^5$ | $10^6$ | $10^7$ | $10^8$ |
|---|---|---|---|---|---|---|
| $\Psi_x^{(x)}$ | 0.146 | 0.606 | 1.444 | 3.194 | 6.049 | 15.336 |
| $x_{max}$ | 0.812 | 0.867 | 0.926 | 0.956 | 0.0272 | 0.176 |
| $y_{max}$ | 0.159 | 0.118 | 0.0702 | 0.0419 | 00969 | 0.0921 |
| $z_{max}$ | 0.465 | 0.414 | 0.364 | 0.315 | 0.745 | 0.877 |
| $\Psi_y^{(y)}$ | 1.130 | 4.997 | 9.966 | 17.863 | 33.041 | 61.932 |
| $x_{max}$ | 0.509 | 0.509 | 0.693 | 0.841 | 0.0844 | 0.0521 |
| $y_{max}$ | 0.500 | 0.653 | 0.812 | 0.912 | 0.0444 | 0.0236 |
| $z_{max}$ | 0.500 | 0.483 | 0.414 | 0.431 | 0.535 | 0.535 |
| $\Psi_z^{(z)}$ | 0.150 | 0.779 | 1.972 | 5.014 | 14.815 | 35.221 |
| $x_{max}$ | 0.517 | 0.586 | 0.535 | 0.154 | 0.133 | 0.960 |
| $y_{max}$ | 0.154 | 0.154 | 0.143 | 0.176 | 0.176 | 0.980 |
| $z_{max}$ | 0.805 | 0.805 | 0.841 | 0.936 | 0.963 | 0.254 |

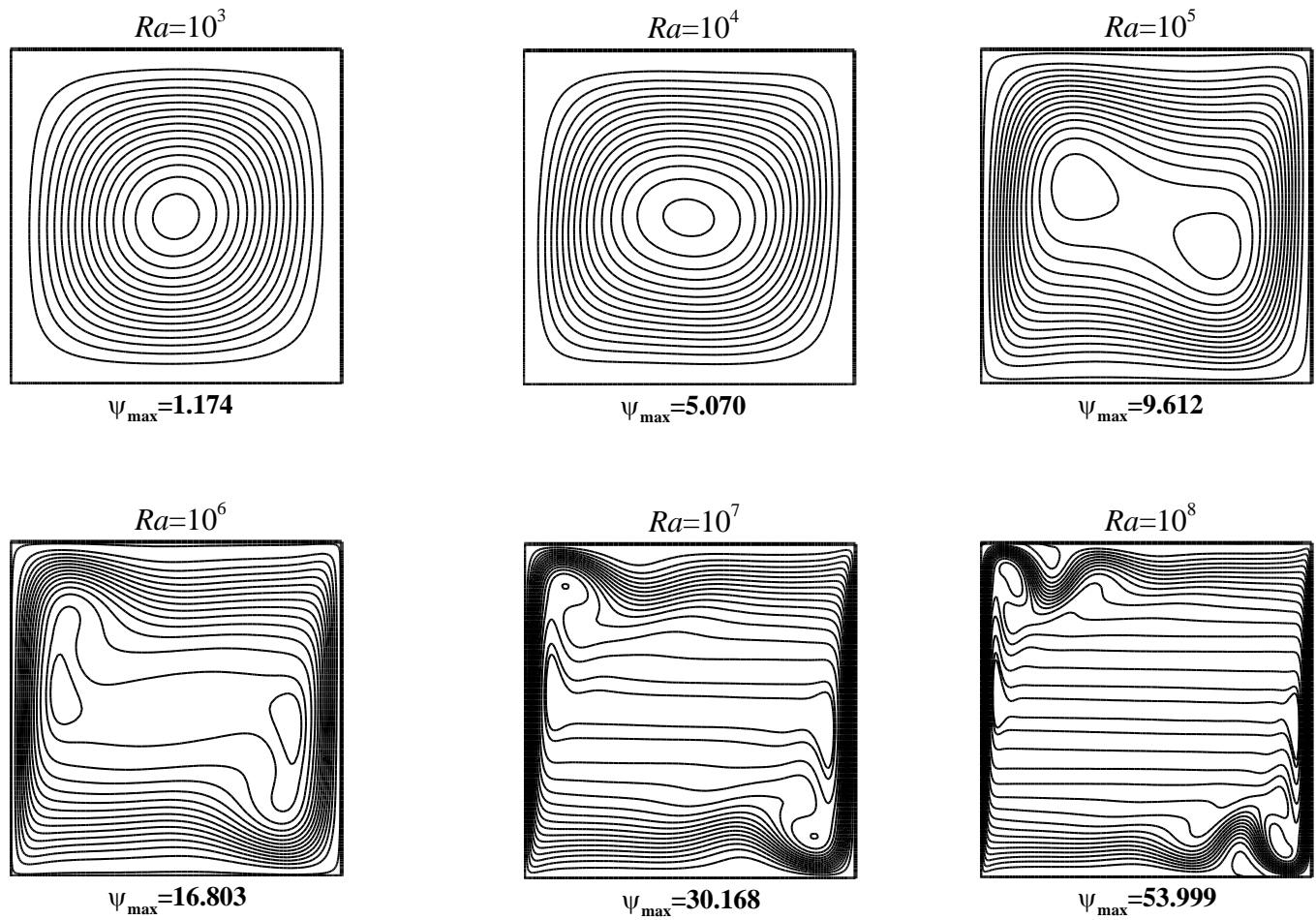

Fig. 1. Streamlines of two-dimensional buoyancy convection flow in a laterally heated square cavity at *Pr*=0.71 and different Rayleigh numbers. The direction of main circulation is clockwise.

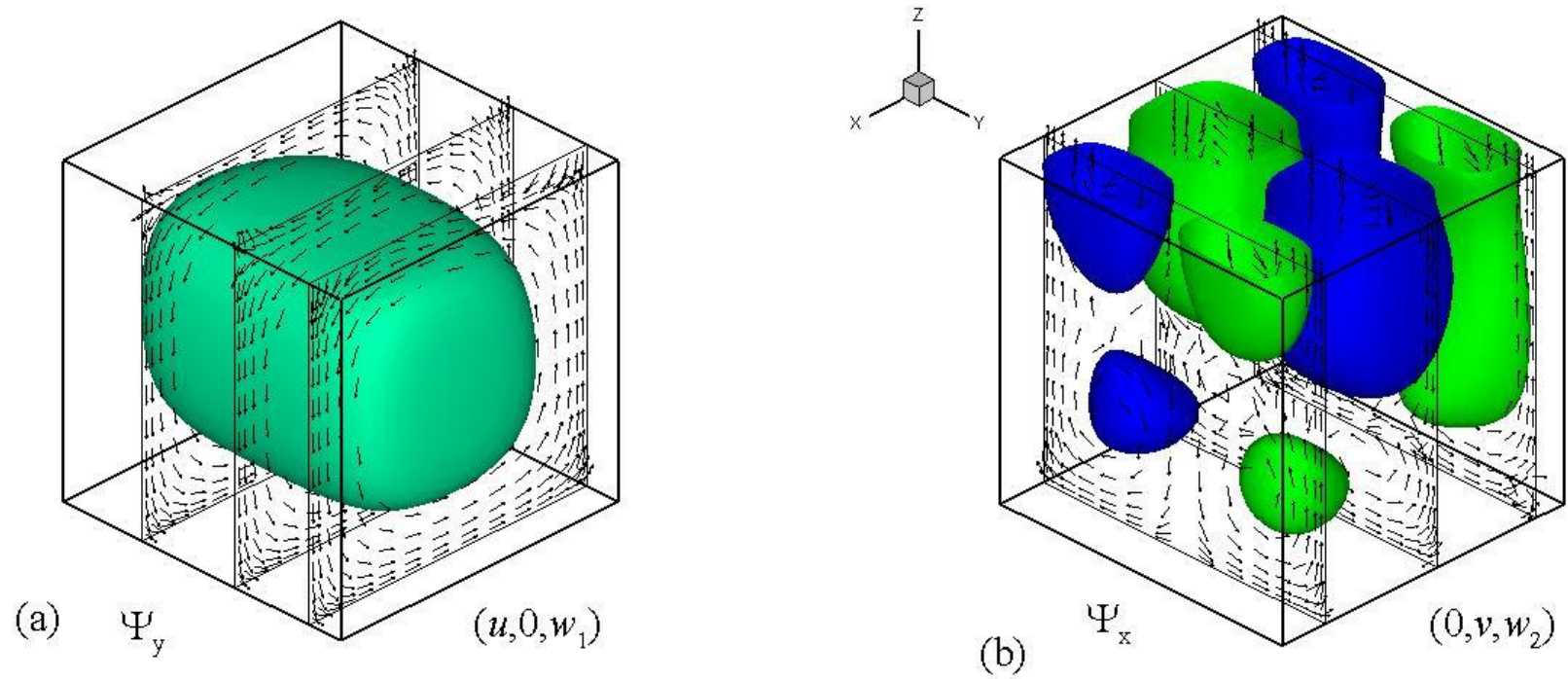

Fig. 2. Calculation for $Ra=10^3$. Vector potentials $\Psi_y$ and $\Psi_x$ defined in Eq. (6) superposed with the vector fields $(u, 0, w_1)$ and $(0, v, w_2)$. (a) $\max \Psi_y = 1.6$, isosurface for $\Psi_y=0.375$; (b) $\max |\Psi_x| = 0.075$, isosurfaces for $\Psi_x = \pm 0.015$.

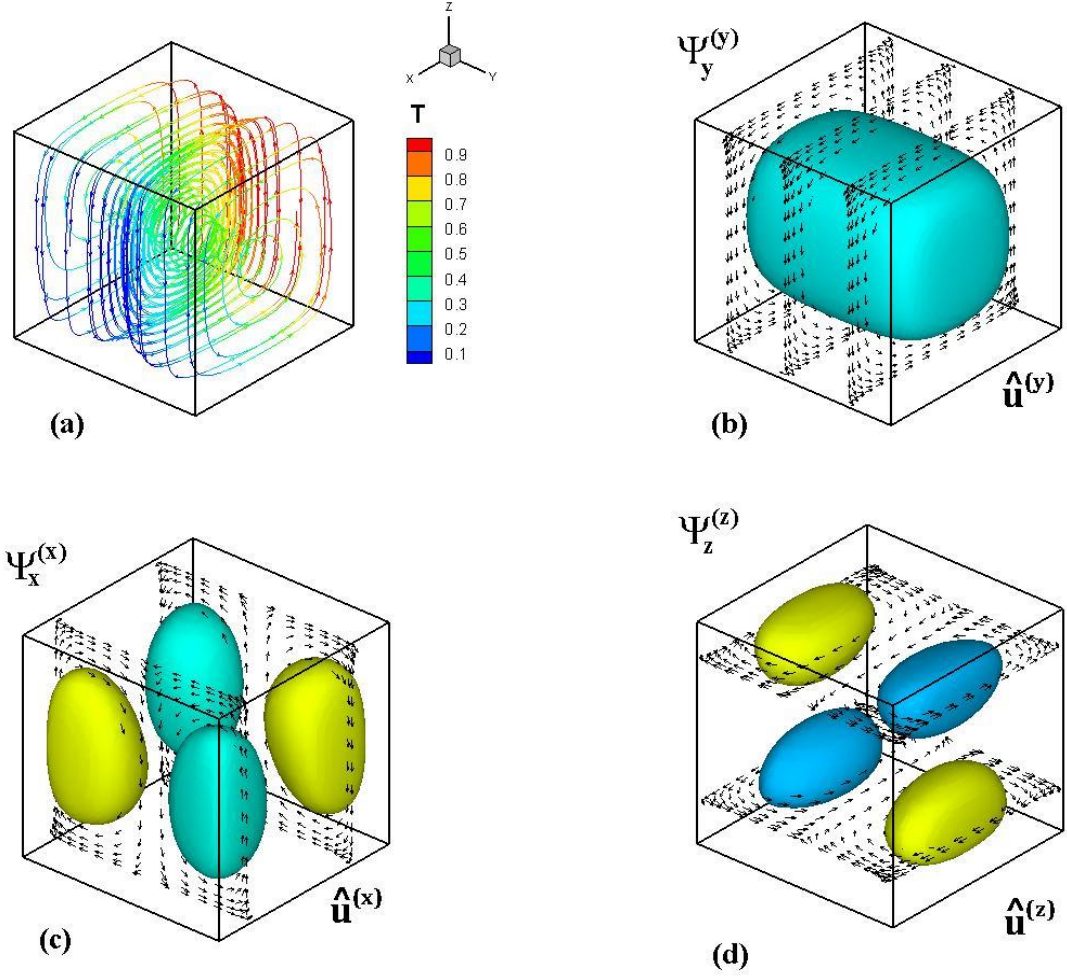

Fig. 3. Visualization of a three-dimensional flow at $Ra=10^3$. (a) Two flow trajectories starting at the points (0.1,0.1,0.1) and (0.9,0.9,0.9). The trajectories are colored due to the temperature values at the points they pass. The temperature color map is shown aside. (b), (c), (d) Isosurfaces of $\Psi_y^{(y)}$, $\Psi_x^{(x)}$ and $\Psi_z^{(z)}$ superimposed with the vector plots of the fields $\hat{\boldsymbol{u}}^{(y)}$, $\hat{\boldsymbol{u}}^{(x)}$ and $\hat{\boldsymbol{u}}^{(z)}$, respectively. The isosurfaces are plotted for $\Psi_y^{(y)} = 0.375$, $\Psi_x^{(x)} = \pm 0.088$, and $\Psi_z^{(z)} = \pm 0.11$.

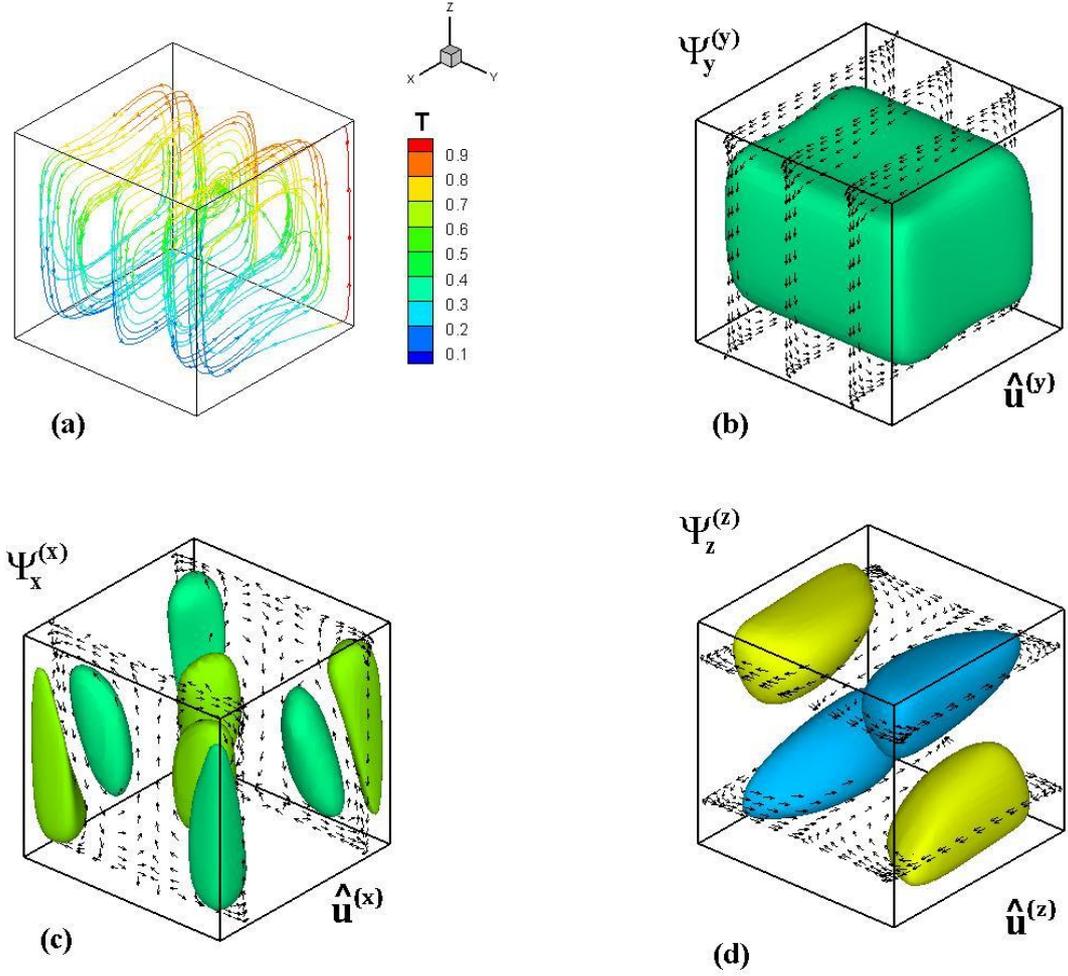

Fig. 4. Visualization of a three-dimensional flow at $Ra=10^5$. (a) Two flow trajectories starting at the points (0.1,0.1,0.1) (0.9,0.9,0.9) and (0.4,0.5,0.5). The trajectories are colored due to the temperature values at the points they pass. The temperature color map is shown aside. (b), (c), (d) Isosurfaces of $\Psi_y^{(y)}$, $\Psi_x^{(x)}$ and $\Psi_z^{(z)}$ superimposed with the vector plots of the fields $\hat{\boldsymbol{u}}^{(y)}, \hat{\boldsymbol{u}}^{(x)}$ and $\hat{\boldsymbol{u}}^{(z)}$, respectively. The isosurfaces are plotted for $\Psi_y^{(y)} = 5.63$, $\Psi_x^{(x)} = \pm 0.64$, and $\Psi_z^{(z)} = \pm 0.27$.

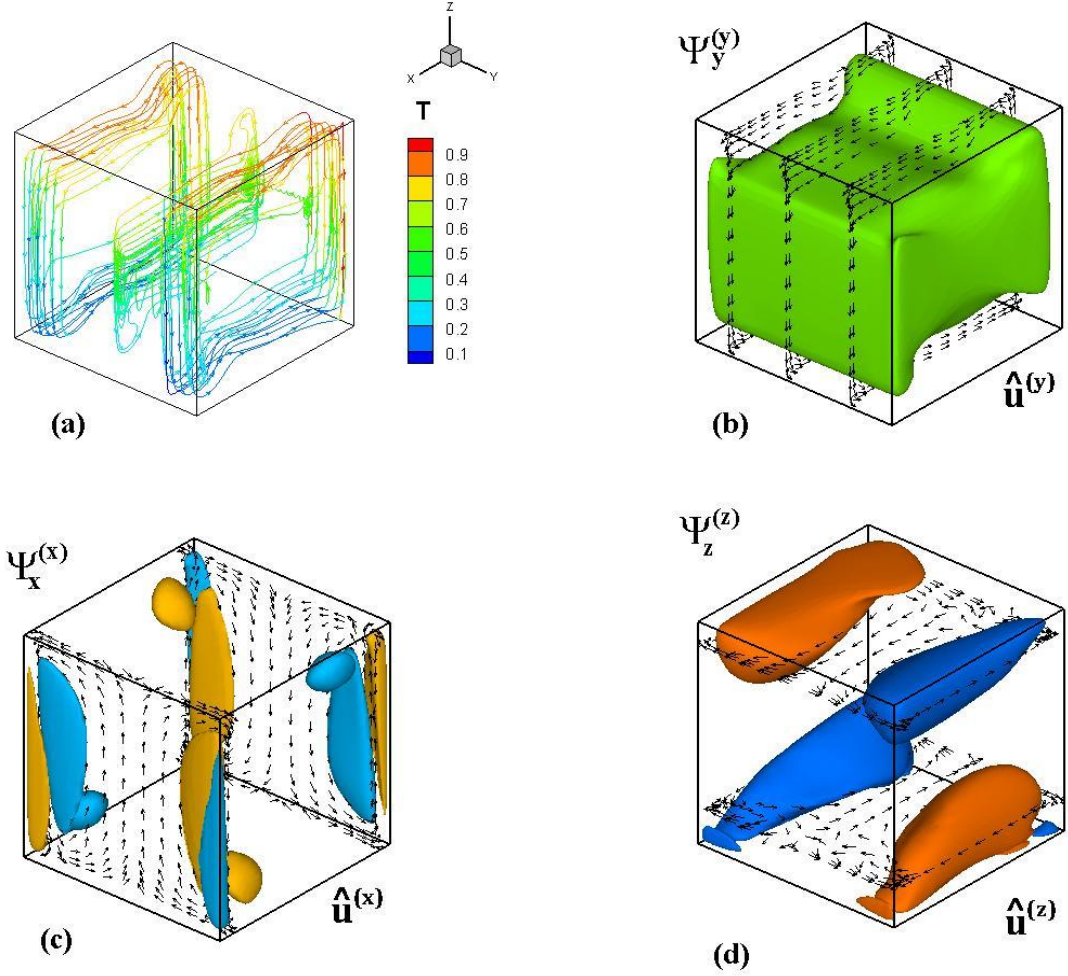

Fig. 5. Visualization of a three-dimensional flow at $Ra=10^7$. (a) Two flow trajectories starting at the points (0.1,0.1,0.1), (0.9,0.9,0.9) and (0.1,0.5,0.5). The trajectories are colored due to the temperature values at the points they pass. The temperature color map is shown aside. (b), (c), (d) Isosurfaces of $\Psi_y^{(y)}$, $\Psi_x^{(x)}$ and $\Psi_z^{(z)}$ superimposed with the vector plots of the fields $\hat{\boldsymbol{u}}^{(y)}, \hat{\boldsymbol{u}}^{(x)}$ and $\hat{\boldsymbol{u}}^{(z)}$, respectively. The isosurfaces are plotted for $\Psi_y^{(y)} = 28.5$, $\Psi_x^{(x)} = \pm 2.6$, and $\Psi_z^{(z)} = \pm 3.0$.

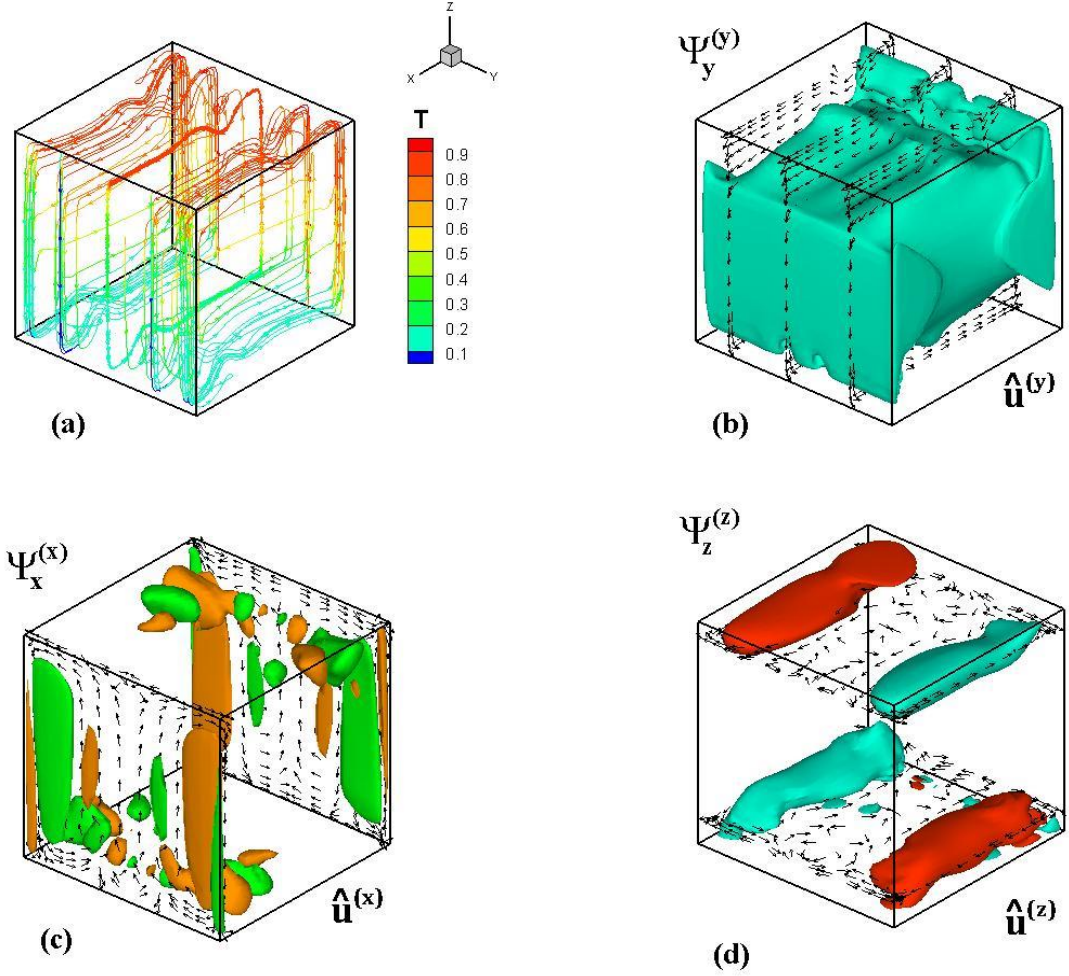

Fig. 6. Visualization of a three-dimensional flow at $Ra=10^8$. (a) Two flow trajectories starting at the points (0.1,0.1,0.1) (0.9,0.9,0.9) and (0.1,0.5,0.5). The trajectories are colored due to the temperature values at the points they pass. The temperature color map is shown aside. (b), (c), (d) Isosurfaces of $\Psi_y^{(y)}$, $\Psi_x^{(x)}$ and $\Psi_z^{(z)}$ superimposed with the vector plots of the fields $\hat{\boldsymbol{u}}^{(y)}, \hat{\boldsymbol{u}}^{(x)}$ and $\hat{\boldsymbol{u}}^{(z)}$, respectively. The isosurfaces are plotted for $\Psi_y^{(y)} = 66.0$, $\Psi_x^{(x)} = \pm 7.0$, and $\Psi_z^{(z)} = \pm 11.0$.

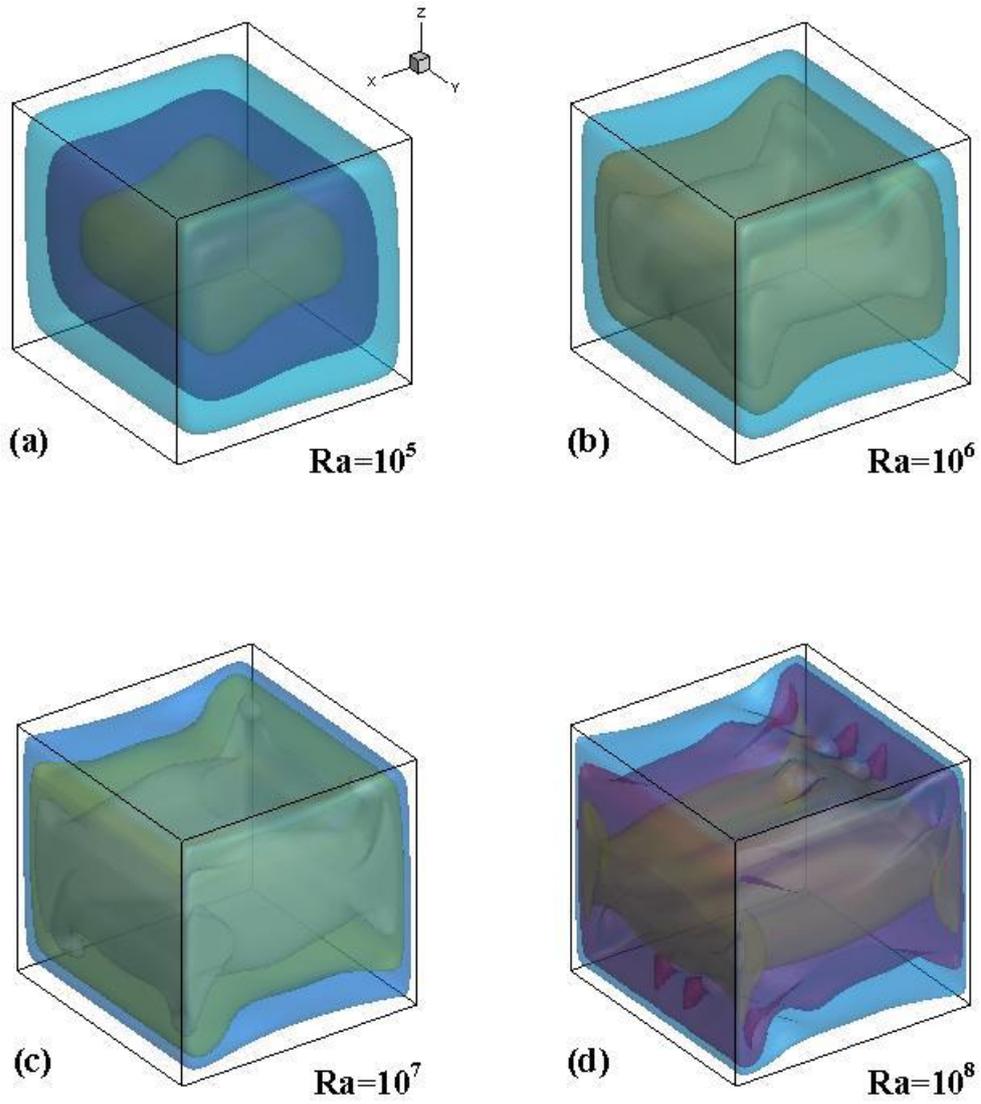

Fig. 7. Isosurfaces of $\Psi_y^{(y)}$ at different Rayleigh numbers. The isosurfaces are plotted at levels (a) 0.75, 5.6, 12.4; (b) 3.6, 15.4, 22.5; (c) 7.5, 24.4, 37.5; (d) 17.8, 47.5, 71.2 .

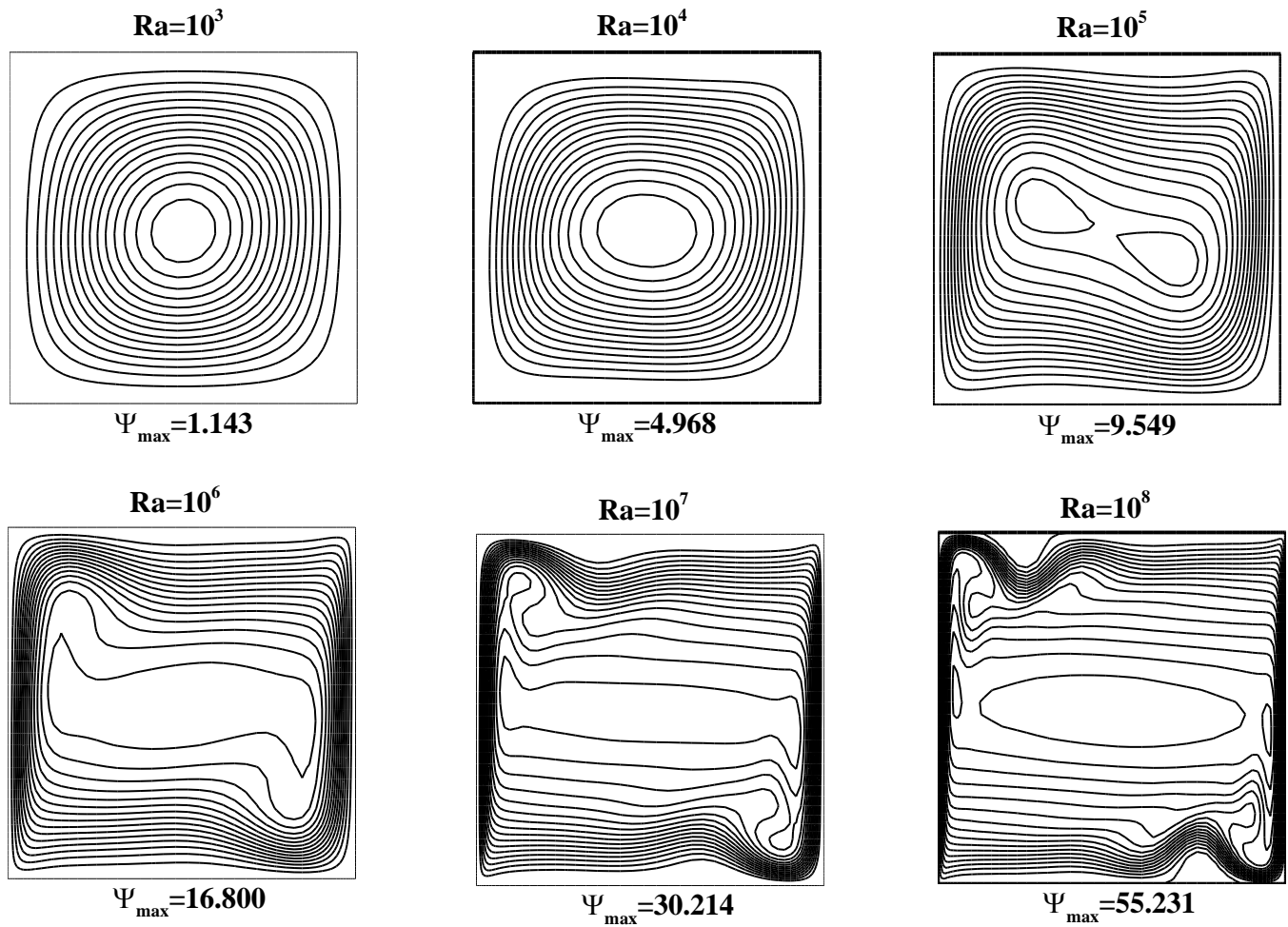

Fig. 8. Isolines of $\Psi_y^{(y)}$ in the midplane $y = 0.5$ at different Rayleigh numbers. The direction of main circulation is clockwise.

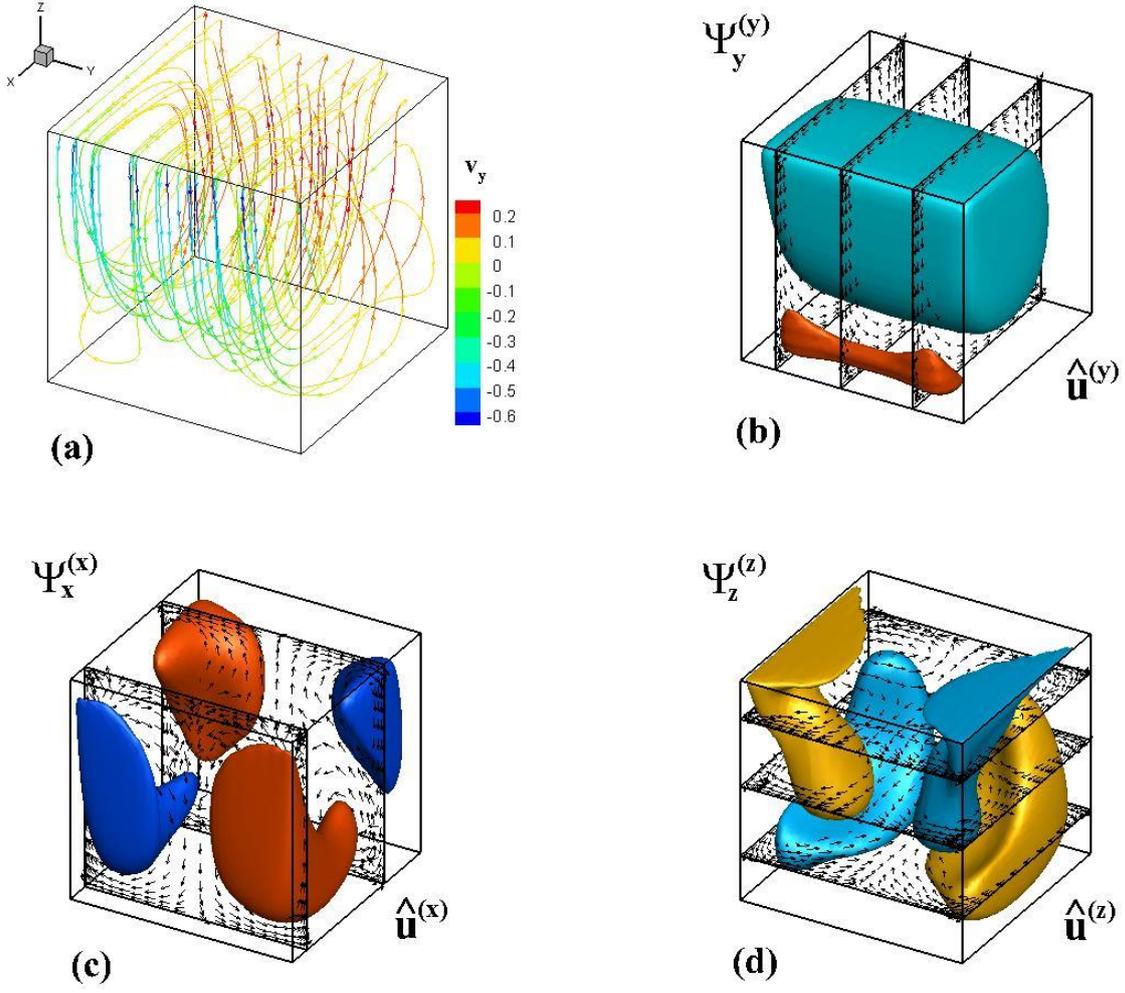

Fig. 9. Visualization of a three-dimensional flow in a lid-driven cubic cavity at $Re=10^3$. (a) Two flow trajectories starting at the points (0.4,0.4,0.9) and (0.6,0.6,0.9). The trajectories are colored due to values of spanwise velocity. (b), (c), (d) Isosurfaces of $\Psi_y^{(y)}$, $\Psi_x^{(x)}$ and $\Psi_z^{(z)}$ superimposed with the vector plots of the fields $\hat{\boldsymbol{u}}^{(y)}$, $\hat{\boldsymbol{u}}^{(x)}$ and $\hat{\boldsymbol{u}}^{(z)}$, respectively. The isosurfaces are plotted for $\Psi_y^{(y)} = -0.0032$ and $+0.0008$; $\Psi_x^{(x)} = -0.0044$, and $+0.0033$; $\Psi_z^{(z)} = \pm 0.0028$.

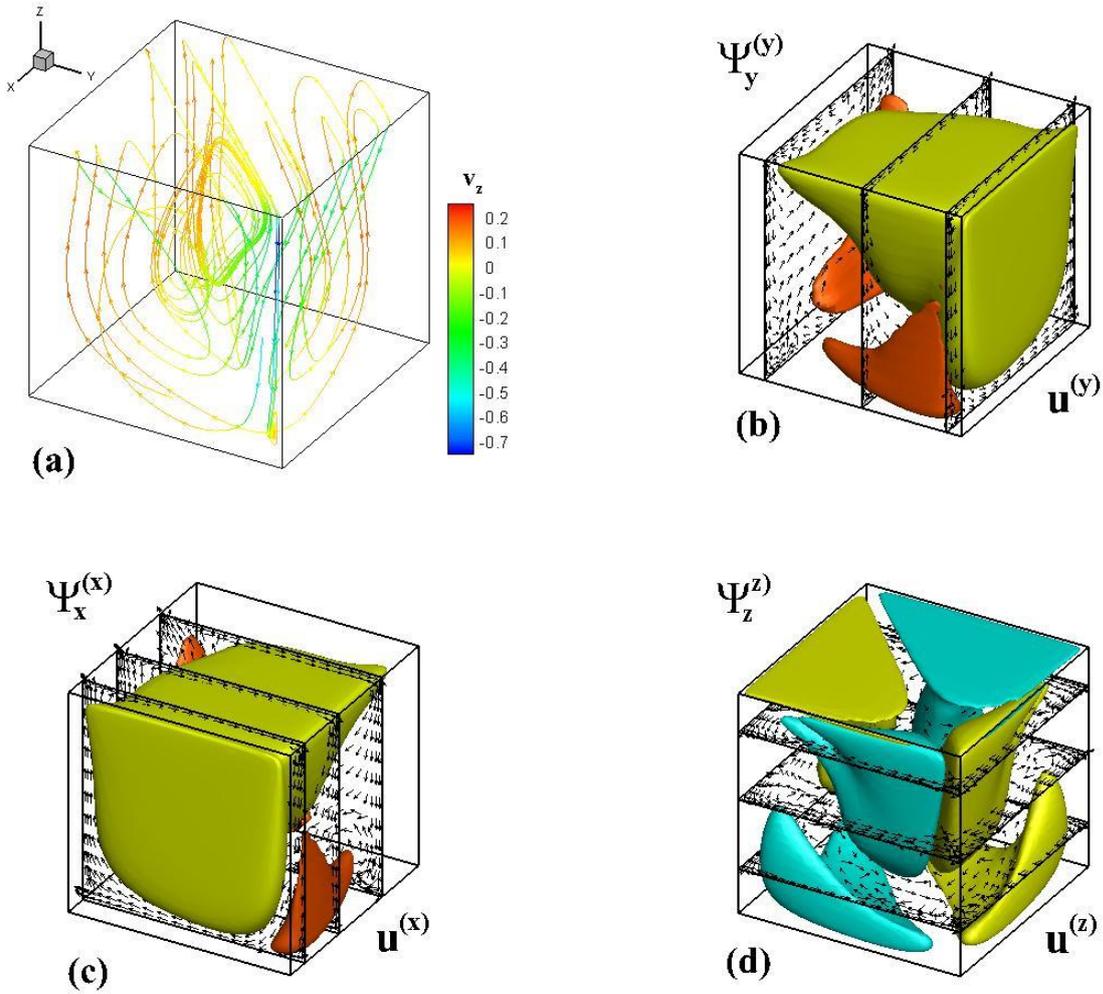

Fig. 10. Visualization of a three-dimensional flow in a lid-driven cubic cavity with a lid moving along a diagonal, at $Re=10^3$. (a) Two flow trajectories starting at the points (0.1,0.1,0.9) and (0.9,0.9,0.9). The trajectories are colored due to values of vertical velocity. (b), (c), (d) Isosurfaces of $\Psi_y^{(y)}$, $\Psi_x^{(x)}$ and $\Psi_z^{(z)}$ superimposed with the vector plots of the fields $\hat{\boldsymbol{u}}^{(y)}, \hat{\boldsymbol{u}}^{(x)}$ and $\hat{\boldsymbol{u}}^{(z)}$, respectively. The isosurfaces are plotted for $\Psi_y^{(y)} = -0.015$ and $+0.0015$; $\Psi_x^{(x)} = -0.015$, and $+0.0015$; $\Psi_z^{(z)} = \pm 0.0024$.

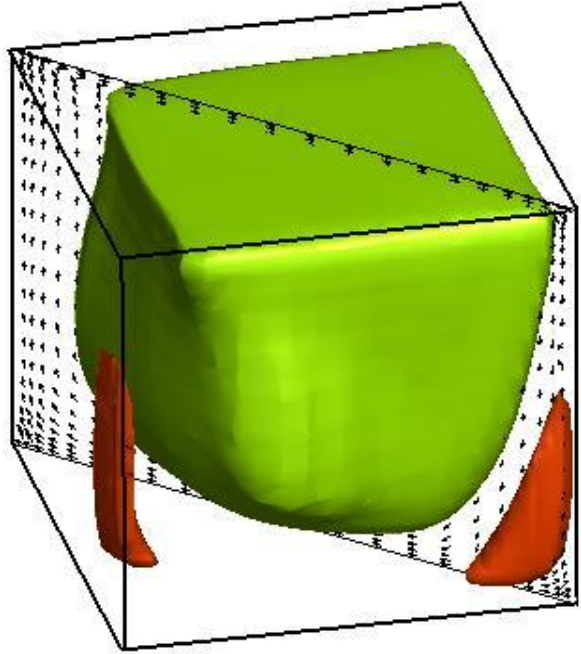

Fig. 11. Visualization of a three-dimensional flow in a lid-driven cubic cavity with a lid moving along a diagonal, at $Re=10^3$. Isosurfaces of vector potential of velocity projection on the diagonal planes, and the vector plot of the corresponding projected velocity field. The isosurfaces are plotted for the levels -0.017 and +0.004, while the minimal and maximal values of the calculated vector potential are -0.083 and +0.012 .